\documentstyle{EuroPhys}  

\newif\ifboo \boofalse

\input epsf

\newcommand{\lan}{\langle}     
\newcommand{\ran}{\rangle}

\newcommand{\be}{\begin{equation}}     
\newcommand{\ee}{\end{equation}}     
\newcommand{\bea}{\begin{eqnarray}}     
\newcommand{\eea}{\end{eqnarray}}     
     
\newcommand{\bef}{\begin{figure}}     
\newcommand{\eef}{\end{figure}}

\def\spose#1{\hbox to 0pt{#1\hss}}      
\def\ltapprox{\mathrel{\spose{\lower 3pt\hbox{$\mathchar"218$}}      
 \raise 2.0pt\hbox{$\mathchar"13C$}}}      
\def\gtapprox{\mathrel{\spose{\lower 3pt\hbox{$\mathchar"218$}}      
 \raise 2.0pt\hbox{$\mathchar"13E$}}}      
\def\inapprox{\mathrel{\spose{\lower 3pt\hbox{$\mathchar"218$}}      
 \raise 2.0pt\hbox{$\mathchar"232$}}}

\begin{document}
\shorttitle{ Baertschiger \& Sylos Labini,  
Initial Conditions in cosmological N-body simulations}  
  
\title{On the problem of initial conditions in 
cosmological N-body simulations}  
\author{Thierry Baertschiger \inst{1} and Francesco Sylos
Labini\inst{1,2} \footnote
{Current Address: LPT, Universit\'e Paris-XI, B\^atiment 211, F-91405
Orsay Cedex, France}, 
 }  
\institute{  
      \inst{1} D\'ept.~de Physique Th\'eorique,  
            Universit\'e de Gen\`eve 24, Quai E. Ansermet,   
            CH-1211 Gen\`eve, Switzerland \\  
      \inst{2} INFM Sezione di Roma1,  Dipartimento di Fisica, 
      Universit\`a di Roma ``La Sapienza'' -  
            P.le A. Moro 2, I-00185 Roma, Italy\\  
        }  
\rec{ }{ }  
\pacs{  
\Pacs{05}{20$-$y}{Statistical Mechanics}  
\Pacs{98}{65$-$r}{Large scale structure of the Universe}  
      }  
\maketitle

\begin{abstract}   
Cosmological N-body simulations aim to calculate 
the {\it non-linear} 
gravitational growth of structures via particle dynamics.
A crucial problem concerns the setting-up of the initial particle 
distribution, as 
standard theories of galaxy formation predict
the properties of the initial continuous density field with small amplitude
correlated Gaussian fluctuations. 
The discretisation of such a field is a 
complex issue and particle fluctuations are 
especially relevant at small scales where non-linear dynamics 
firstly takes place. In general, most of the procedures
which may discretise a continuous field, gives rise to Poisson noise, which
would then dominate the non-linear small-scales  
dynamics due to nearest-neighbours interactions.
In order to avoid such a noise, and to consider the dynamics as due only
to  large scale (smooth) fluctuations, an ad-hoc method 
(lattice or glassy system plus
correlated displacements) has been 
introduced and used in  cosmological simulations.
We show that such a  method gives rise to a particle distribution
which {\it does not have } any of the correlation properties
of the theoretical continuous density field. This is because 
discreteness effects, different from Poisson noise
but nevertheless very important, determine  particle fluctuations
at any scale, making it completely different from the original
continuous field. We  conclude 
that discreteness effects play a central role in the 
non-linear evolution of N-body simulations. 
\end{abstract}

The purpose of cosmological N-body simulations is to calculate the
{\it non-linear} 
growth of structures in the universe by following individual particles
trajectories under the action of their mutual gravity
\cite{efst85,white93,jenkins98}. These particles are not galaxies
but are meant to represent some sorts of collisionless clouds of
elementary dark matter particles. 
In order to make them move, one must calculate the force
acting on each of them due to all the others. In general one may find
several algorithms  which speed up the $N^2$ sum necessary to
compute the force on each particle. In this paper we study simulations
from the Virgo project \cite{jenkins98} which are done with an
``adaptative P$^3$M'' scheme \cite{couch91,pad}.  This scheme is a 
combination  
of a PM (i.e. meshed based scheme) and a PP (i.e. particle-particle 
based scheme).
Briefly, this
consists of assigning the particle mass on a mesh so as to obtain a
density field on the mesh itself. One can then calculate the gravitational
potential on that mesh by solving the Poisson equation, and finally
the potential at the location of a particle is determined by an
interpolation of the values on the mesh. For a better resolution, a
true calculation of the force due to nearby particles is done. In
highly clustered regions, since this last calculation can be quite
long, additional higher resolution meshes are placed. The force used
is not a pure $r^{-2}$ one:  instead one smoothes it at small $r$
by choosing for instance a force proportional to
$(r^2+\epsilon^2)^{-1}$ in order to avoid ``collisions'' between close
particles. This brings us to an important hypothesis sometimes made 
in cosmological N-body simulations: with a softened force and a proper
choice of the softening parameter $\epsilon$, the evolution of the N-bodies
should be the same as the evolution of a continuous density field
(made of a huge number of  particles behaving like a
fluid) under its own gravity. With this in mind, one interprets
collisions, or more correctly strong scattering, between nearby
particles in the simulation as 
due to the discretisation of
the density field and therefore artificial \cite{melott,splinter}. 
However, it is
important to note that, in order to satisfy the above hypothesis, 
$\epsilon$ should be at least as large as the mean
inter-particle distance $\langle \Lambda \rangle$
\cite{melott,splinter}:  this is not always
the case as for instance in the Virgo project 
where $\epsilon = 0.036 Mpc/h$ \cite{jenkins98} 
and $\langle \Lambda \rangle  \approx 0.5 Mpc/h$ (see below) .

The philosophy behind cosmological N-body simulations is to reproduce
galaxy structures that we see today through redshift surveys by choosing
some good parameters (cosmological and numerical) for the evolution of
the system and especially fine-tuned initial conditions 
(properly normalised to the primordial fluctuations) given by some
theories. This last point is important because one is not
interested in some general asymptotic behaviors or quasi stationary
states independent of the initial conditions, but in the state of the
system after a relatively short time compared with the dynamical time of 
the simulation.

Initial conditions (hereafter IC) 
are created from  theories typically based on
inflation and the subsequent evolution of matter, 
which are able to predict
the properties of a continuous density field with correlated
fluctuations \cite{bond84}.  
Such a density field is usually 
Gaussian, and
hence all the statistical properties are contained in the two-point
correlation function (hereafter CF) or its Fourier transform, the power
spectrum (hereafter PS). 
These two functions depend on the kind of dark matter which
is the relevant one on large scales: hot (relativistic), cold
(non-relativistic) or warm (a mixture of the two). However
the main point for
what concerns us is
that a continuous and smooth density field 
with correlated density fluctuations 
is  given as IC.  If one wants to study the time evolution
of this field with N-body simulations based on particle
dynamics, it is then necessary to discretise the field. This means
that one has to create a particle  distribution which 
is representative 
of the continuous density field and to control any finite size 
effect. 
Note that 
a crucial point with respect to structures formation, is that non-linearities
firstly develop at small scales where discreteness effects are important.
Firstly,  we briefly
explain the standard way of carrying out the discretisation  
and  then we analyse 
some of the real space statistical properties of these
distributions, restricting ourself to a Virgo Standard Cold Dark Matter 
(hereafter CDM)   simulation
with $256^3$ particles \cite{jenkins98}.  The
questions we address are: ~In which range of scales the initial
particle distribution used in the standard cosmological simulations is
representative of CDM-like 
density fluctuations? What drives the non-linear dynamics
of structures formation:  
the discreteness effects due to the short NN 
interaction
or the large scale (continuous) distribution of density fluctuations?

%
%

The standard ad-hoc procedure for setting up IC is described in 
\cite{efst85,white93,baugh95,jenkins98}.  For galaxy structures formation
problems, the IC generation splits into two parts.  The first is to
set up a ``uniform'' distribution of particles, which can represent
the unperturbed universe. The second is to impose density
fluctuations with the desired  characteristics.  
In this context, one faces the
general problem of how to discretise a continuous density
field.  There are different procedures (e.g.  random sampling,
threshold sampling, etc.)  which can be chosen and they result in
different point distributions. Clearly one should have some physical
reasons to choose one or another since any procedure introduces some
discreteness effects, like Poisson noise, which  could then 
play the main role in the non-linear dynamics of the system. 
For instance, in a Poisson
distribution
the dominant part of the gravitational force acting on an average
particle is due to its NN \cite{chandra43,gslp99,bottaccio01}.  
If a
simulation is run from such IC the fluctuations grow rapidly into
non-linear objects at small scales. 
This happens because a Poisson
distribution is {\it statistically isotropic} 
only on scales larger than the average distance between
NN \cite{chandra43,gslp99} and the three-point correlation
function is non-zero at these small scales: such a situation
avoids the perfect cancelation of the net gravitational
force due to close particles.
Instead,  in cosmological N-body simulations,
one would like to simulate 
a system where the main contribution 
for non-linear structures formation problem,
is due to the large scale
distribution of the
other particles and 
{\it not to local  NN interactions} \cite{white93}. 

To overcome the fact that a Poisson distribution is only
``statistically'' isotropic 
the most widely used solution to this problem has been to represent
the {\it unperturbed universe} by a {\it regular cubic grid of particles}
\cite{efst85,white93,baugh95}. An infinite lattice, or a lattice
with periodic boundary conditions, is ``gravitationally
stable'' because of symmetry reasons.
However a lattice is a distribution with fluctuations at all scales
{\it and} non-trivial correlations.  One can show that the
unconditional variance in spheres 
decays as $\sigma^2(r) \equiv \langle \Delta
N(r)^2 \rangle / \langle N(r) \rangle^2
\sim r^{-4}$ (where  $\langle
N(r)\rangle \sim r^3$ 
is the average number of points in a {\it randomly chosen} 
sphere of radius $r$ \cite{gsl01}) 
and this is due to the fact that a lattice is a
strongly ordered and correlated system at all scales. The 
two-point
CF is such that $\xi(\vec{r}_1,\vec{r}_2) =
\xi(\vec{r}_1 -\vec{r}_2) \neq \xi(|\vec{r}_1 -\vec{r}_2|)$ because
it is not invariant for space rotation \cite{hz}: A lattice breaks space
isotropy.
Moreover,
the grid-like system has the disadvantage to introduce a strong
characteristic length on small scales - the grid spacing - and it
leads to strongly preferred directions on all scales.

For the latter reasons, a second way to generate an ``uniform
background'' is by means of the following procedure.  
One starts from a Poisson distribution and then the N-body integrator is
used with a {\it repulsive} gravitational force in such a way that,
after the simulation is evolved for a sufficiently long time, the
particles settle down to a {\it glass-like configuration} in which
the force on each particle is very close to zero, i.e. a global
stable configuration is found  which has no preferred directions
\cite{white93}.  
The resulting distribution is very
isotropic but it is still characterized by long-range order of the
same kind of a lattice.  As for the lattice the main characteristic
of such a distribution is the presence of an {\it excluded volume}:
two particles cannot lie at a distance smaller than a certain fixed
length scale \cite{hz}. In the lattice this scale is the grid space, for a
glass such a distance depends on the number of points one has
distributed in a given volume.  The fact that a lattice is ordered
is due to the existence of the deterministic small scale distance.
The unconditional variance scales as $\sigma^2(r) \sim r^{-\alpha}$
where $3< \alpha \le 4$, and it is again a strongly correlated
system \cite{hz}. 
Its two-point 
CF  depends on the detailed
procedures used to generate the glass distribution. However it is
possible to show that this class of distribution has $P(k) \sim k^a$
(with $a>0$) for $k \rightarrow 0$ and hence $P(0)=0$ like the
Harrison-Zeldovich PS  \cite{hz}.


Given a ``suitably unperturbed'' particle distribution, any desired
{\it linear} fluctuation distribution can be in
principle generated using {\it
Zeldovich approximation} \cite{efst85}.  
Basically, one assumes
that the initial background is uniform with average 
density $\rho(\vec{r})=\rho_0$,
without fluctuations, and then one applies a displacements field
$\vec{u}(\vec{r})$. Because of the conservation of total mass after
the displacements, one can apply a continuity equation which gives
\be
\label{aq1}
\rho(\vec{r}) -\rho_0 + \vec{\nabla} 
\cdot \Bigl( \rho_0 \vec{u}(\vec{r}) \Bigr)= 0\;.
\ee
If statistical homogeneity and isotropy of the displacements
field $\vec{u}(\vec{x})$ is assumed, then on going to 
Fourier space and considering the
expectation value of the square modulus of the density fluctuations
(the PS),
Eq.\ref{aq1} leads to
$P(k) = 
\langle \delta_k^2 \rangle = k^2 \langle | \vec{u}(\vec{k})|^2 \rangle 
= k^2 P_u(k).$ 
As any  
PS  cannot diverge faster than $k^{-3}$ for $k \rightarrow 0$,
we have that
$\lim_{k \rightarrow 0} P(k)\sim k^n \;\; \mbox{with} \;\;n> -1.$
This means that, for instance,  
one cannot create a density fluctuation field with
$P(k)$ which diverges as $k^n$ and $n \le - 1$, or with CF
 which goes as $r^{-\alpha}$ for $r\rightarrow \infty$ and
$\alpha \le 2$ only by applying a displacement field to a uniform
background.  Moreover 
one does not start from
a continuous field with no fluctuations, 
but with a particle distribution, which has its
own fluctuations and correlations (which in the lattice's and glass's
cases are long-range). 
In such a situation one has to
ensure that the correlations among density fluctuations implemented by
the displacements field are larger than the intrinsic fluctuations
of the original particle distribution, and that the large-scale fluctuations
dominate the non-linear small scales clustering instead of 
nearest-particle interactions.
Only if one considers
very large scales, and/or a displacements field which introduces
correlations which are larger than the intrinsic one of the original
distribution, one can recover the PS as in Eq.\ref{aq1}.  
Otherwise
in a certain range of small enough scales,  
the point distribution is dominated
by discreteness effects, which in this context
can be seen as finite-size effects and which 
are important for what concerns the small-scale non-linear 
structures formation.

In order to clarify this question, we have 
checked numerically the kind of  
correlations introduced
in the system with the Zeldovich approximation: 
We have analysed a CDM simulation with $\Omega_0=1$
\cite{jenkins98} as an example, but our result are generally valid
for any other particular model chosen, as they involve {\it the same method
for setting-up initial conditions}.  Note that we have analysed the
statistical quantities for the entire simulation ($N=256^3$ particles)
and {\it not} for a sub-sample of it, in order to avoid introducing any
kind of sampling noise and we have used periodic boundary conditions.  
The pre-initial particle  distribution (which would represent the 
``uniform background" as previously discussed) 
consists of replicas of a $N=10^6$ particles 
glass distribution generated by the
procedure discussed above.  As they are generated with periodic
boundary conditions they can be tiled to make larger distributions.
After the Zeldovich displacements one obtains the IC 
which correspond to a redshift $z=50$.
(In general the particles are very little displaced with respect 
to their initial positions: that is the average distance between
NN remains unperturbed.)
 
It is worth noticing that in general \cite{efst85,baugh95} the
correlation properties of the initial particle distribution of N-body
simulations have been discussed through the analysis of the PS of the
density fluctuations.
The PS is the Fourier transform of the real
space CF $\xi(r)$ and hence it contains the same
information, clearly if finite size effects and statistical noise {\it have
been properly taken into account}.  
For the comparison with theoretical CDM models, which usually 
give the PS of
density fluctuations \cite{bond84},  we have computed the real
space properties:
$\xi(r)=
\frac{1}{2\pi^2} \int_0^{\infty} P(k) \frac{\sin(kr)}{kr} \, k^2 \, dk$   
and   
$\sigma^2(R) = \frac{1}{2\pi^2} 
\int_0^{\infty}  \tilde W(k,R)^2 P(k) \,  d^3k$
where $W(k,R)$ is the Fourier transform of the sample's window
function (a real space sphere in our case - 
which is defined to be zero outside the sample and one
inside, and its integral over all space is one).
In the simulation, the CF is computed  
by direct pair counting (we have used the estimator introduced
by \cite{dp83}), and $\sigma^2(r)$  by
distributing $N_s$ spheres of varying radius with random centers and
computing the number fluctuation.  For the latter we
have used $N_s=2\cdot 10^4$ spheres randomly distributed in the
simulation volume and we define our estimator as
\be
\sigma_E^2(r) = \frac{1}{\lan N(r) \ran^2} 
\sum_{i=1}^{N_s} \frac{(N_i(r)  - \lan N(r) \ran)^2}{N_s-1}
\ee
where $N_i(r)$ is the number of points contained in the $i^{th}$ sphere of
radius $r$ and $\lan N(r) \ran$ is its average.  
One may see in Fig.\ref{fig1} 
that for a CDM model, the  
$\sigma^2(r)$ is constant up 
to a scale $r_c \approx 0.06$  
(normalized to 
the simulation box $L=239.5 Mpc/h$),
which is fixed by the turn-over scale of the PS
\cite{hz}, and then it decays as $r^{-4}$. 
(The normalization has been performed  by 
requiring that $\sigma(r= 8Mpc/h) = 0.51$ as 
in \cite{jenkins98}: we note however 
that a different normalization does not change 
our main results which concerns the {\it functional behaviour} of the 
real space properties with scale.)
For the glass $\sigma^2(r) \sim r^{-4}$ at any scale, as expected, 
while for the IC $\sigma^2(r)$  decays as $r^{-4}$ up to
$\langle  \Lambda \rangle$ and then it decays slower as 
$r^{-1.6}$: this change of slope is the effect
of the displacements field. 
In Fig.\ref{fig2} we show the results for
the two-point CF. 
For the CDM model used here \cite{bond84} 
$\xi(r) \sim const.$ up to $r_c$ and then after crossing 
zero,  it goes as $\xi(r) \sim -r^{-4}$ \cite{hz}. We find instead that
$\xi(r)$ for the IC and the glass are rather similar at small scales
and they oscillate around zero: the peaks due to the 
first, second and third NN are  clearly visible. 
Note that the average distance between NN is $\langle \Lambda \rangle
= 0.003$ (in normalized units) for both the glass-like distribution
and the IC and it corresponds to the first peak of the conditional
average 
density $\sim \langle n \rangle (1+\xi(r))$ 
(see Insert Panel of Fig.2).  
At large scales there 
is a slight difference between the two, which changes the 
behaviour of $\sigma^2(r)$. However in any range of scales 
there is no  agreement between the $\xi(r)$ and $\sigma^2(r)$ 
of CDM and IC. This is especially  relevant for the nature
of  fluctuations in IC and CDM: for the latter one should see 
an over-density up to a scale $r_c$ followed by a peculiar 
under-density (with $\xi(r) \sim -r^{-4} $)
while for the first we find that they are a sequence of over-densities
and under-densities which results in an oscillating 
$\xi(r)$.


Let us now discuss the second point of this paper.
A lattice
or a glass are gravitationally unstable with respect to 
small perturbations of their configuration. 
For this reason an interesting question
concerns what drives the non-linear dynamics in these simulations.
Once the Zeldovich displacements have been implemented, the
new configuration does not have the perfect symmetry
of the pre-initial distribution. Hence the question
concerns: 
are  discreteness effects 
dominant for the small-scales
non-linear dynamics with respect to  the force due to 
large scales smooth  fluctuations ?
In order to study this point, we have computed (see Fig.\ref{force}) 
the behaviour
of the mean gravitational force on a particle due to all the particles
contained in the  sphere of radius $r$ around,   
as a function of $r$.
Firstly, we note that the force increases of a factor 10 approximately
between the glass and the IC. In the case of the latter, one can see
that the force due to the first shell is almost compensated  
by the second one
and starts to grow at least until the sphere reaches the size of the
box. Furthermore the force due to the first shells is one third of the
force at large distance and fluctuations are of the same order than
the average. One can therefore say that the contribution from the
NN  is not as important as in the Poisson case because
the force doesn't reach its asymptotic value at scales $\sim 2 \langle
\Lambda \rangle $ \cite{gslp99}. 
The average force increases with scale up to
the box's size; however the fluctuations at all scales are 
large enough to conclude that small-scales discreteness effects are not 
negligible with respect to the large scale
contribution.

This last analysis would be totally useless without taking into
account the initial velocity of the particles. Indeed if the particles
were too fast they would not be trapped by nearby particles and
discreteness would  only make particles trajectories less regular.
However, we expect the particles to be slow since we 
study CDM. In order to make this more quantitative, we did the
following test: calculate the force on a particle due to particles in
a sphere around, calculate the velocity of the particle in a direction
perpendicular to the force and finally compare the time needed to
travel a certain distance with this velocity but without any force
($t_{v}$) and the time to travel the same distance with the force but
without any initial velocity ($t_{f}$). Our result is that the
velocities are indeed small. For instance if we take $\langle\Lambda
\rangle/10$ for the distance to travel, one has that $t_{v}$ goes from
$\sim 3t_{f}$ to $\sim 10 t_{f}$ when the radius of the sphere for the
force goes from $0.004$ to $0.4$. This shows that structures should be
able to form even at the smallest scales $r\sim \langle\Lambda \rangle$ in
the simulation. 
Indeed,
at a distance $\langle\Lambda \rangle/10$ from its initial position, 
the force from its NN  could be already driving a
particle, instead of its initial velocity.
Discreteness effects are therefore important
for what concerns the non-linear growth of structures 
at small scales and  this is  the relevant range of scales
where one wants to study strongly 
or weakly  ($\xi(r) \gtapprox 0.1$) non-linear regimes
\cite{jenkins98}. (see \cite{thierry2} for a more
complete discussion of this point and of the whole
time evolution of the simulations).

Let us briefly summarize the discussion.
The idea of the standard procedure to set-up 
IC in cosmological N-body simulations,
 is that one starts with a ``uniform 
distribution'' 
which, once discretised, can be a lattice or a glassy system. 
Then 
one gives correlated displacements to the particles 
in such a way that one gets a system which should behave like a continuous
fluid with correlated density fluctuations which  
have the desired (i.e. representative of the continuous field) 
correlation properties. 
However, 
that one considers  point distributions
which introduce discreteness effects due to their 
intrinsic fluctuations and correlations.
We have addressed the problem whether 
these discreteness effects due the imprint 
of the (correlated) fluctuations of the 
pre-initial point distribution 
are strong enough to super-seed 
the given correlations. Indeed, we have shown that this is the case
and we have pointed out that the standard ad-hoc procedure
used to create N-body IC
{\it does not give rise to 
the desired CDM-like correlation of density fluctuations}.
This implies that the small-scales non-linear 
dynamical evolution
of the system is driven by fluctuations
which arise from the particular ad-hoc procedure 
used to discretise the field.
{\it In this context, these fluctuations 
can be seen as finite size effects, and 
are completely different from CDM-like fluctuations.}
Moreover  we have studied  the behaviour of the 
gravitational force acting on an average particle 
with the result that the force due to the nearest 
particles is highly fluctuating and gives 
a contribution of the order of the large scale one.
For this reason we conclude that  discreteness 
 effects can be significant relative 
to the smooth distribution 
and that 
they therefore play
an important role in the small-scales non-linear 
evolution of cosmological N-body simulations.
Finally we stress that most of procedures used to discretise 
a continuous field (e.g. threshold or random  sampling)
unavoidably introduces Poisson noise. In such a situation,
due to the strong NN interactions, the large scale 
fluctuations play an even smaller role in the non-linear dynamics
which would be dominated mainly by particle-particle interactions. 

{\bf Acknowledgments}   
We thank  Y. Baryshev, M. Bottaccio, T. Buchert, R. Durrer, A. Gabrielli,
A. Jenkins, M. Joyce,  D. Pfenniger, 
L. Pietronero and D. Steer for 
useful 
discussions, comments and suggestions.
This work is
supported by the      
EC TMR Network  "Fractal structures and  self-organization"       
\mbox{ERBFMRXCT980183} and by the Swiss NSF.


\newpage 

\bef   
 \caption {\label{fig1}
Behaviour of $\sigma^2(r)$ for the pre-initial glass distribution,
for the initial condition of the simulation ($z=50$) and for the theoretical 
expectation for a CDM model. 
For the glass configuration $\sigma^2(r) \sim r^{-4}$, while for the IC 
it deviates from this behavior at $r\sim \langle \Lambda \rangle$ 
beyond which  it behaves as $r^{-1.6}$. For the CDM model
instead $\sigma^2(r) \sim const.$ up to $r\sim r_c \sim 0.1$ and then it decays
as $r^{-4}$. (For the x-axis
we have used normalized units to the box size
$L=239.5 Mpc/h$). Note that there is no agreement
at any scale, between IC and CDM.
}
\eef

\bef  
\caption{\label{fig2}
The behaviour of $\xi(r)$ for the same distributions of the previous
figure
(for the x-axis
we have used normalized units to the box size
$L=239.5 Mpc/h$).
For the glass and IC $\xi(r)$ is oscillating around zero.
For the theoretical 
expectation for a CDM 
one should find $\xi(r) \sim const.>0$ for $r \le r_c  \sim 0.1$
 and then 
$\xi(r) \sim -r^{-4}$ at larger scales.
Note that the nature of the fluctuations in the IC and CDM
distribution is rather different: for the first one has
a sequence of over-densities and under-densities (i.e. an oscillating
$\xi(r)$) while in the second case one would expect 
an over-density $\xi(r) >0$
 followed by an under-density $\xi(r) < 0$. In the insert panel
it is shown the behaviour of the conditional  density $\sim [\xi(r) +1]$
which makes clear the oscillatory nature of $\xi(r)$. 
}
\eef

\bef  
\caption{\label{force}
The behaviour of the average total force and its variance 
as a function of distance (for the x-axis
we have used normalized units to the box size
$L=239.5 Mpc/h$)
for the pre-initial (glass) and initial particle distribution.
For the glass configuration the force is very small
and almost zero.
For the IC, 
one may note that the NN gives a very fluctuating contribution
to the total force of the order of the asymptotic one.
Finally, note that the force does not converge at the box size.
}
\eef

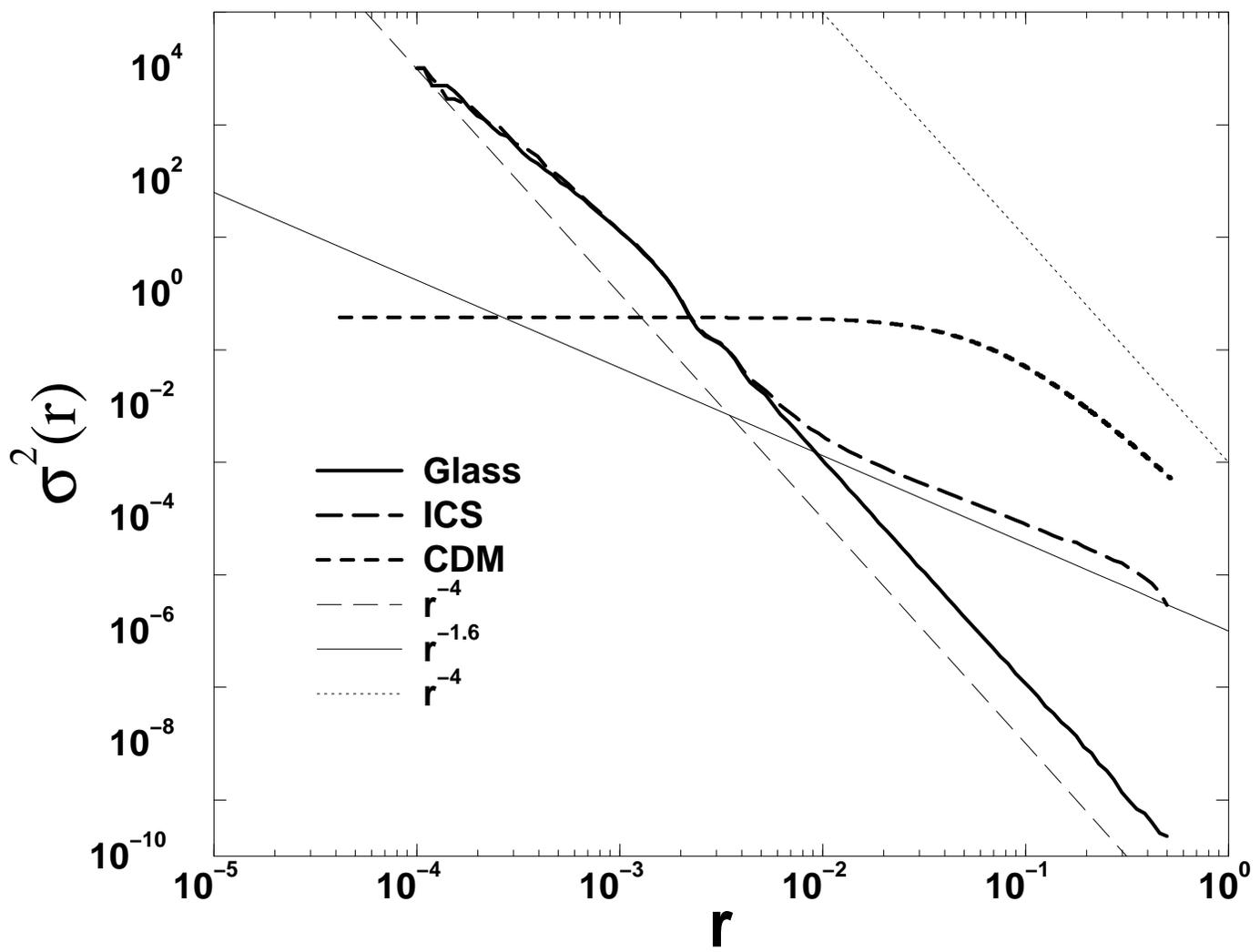
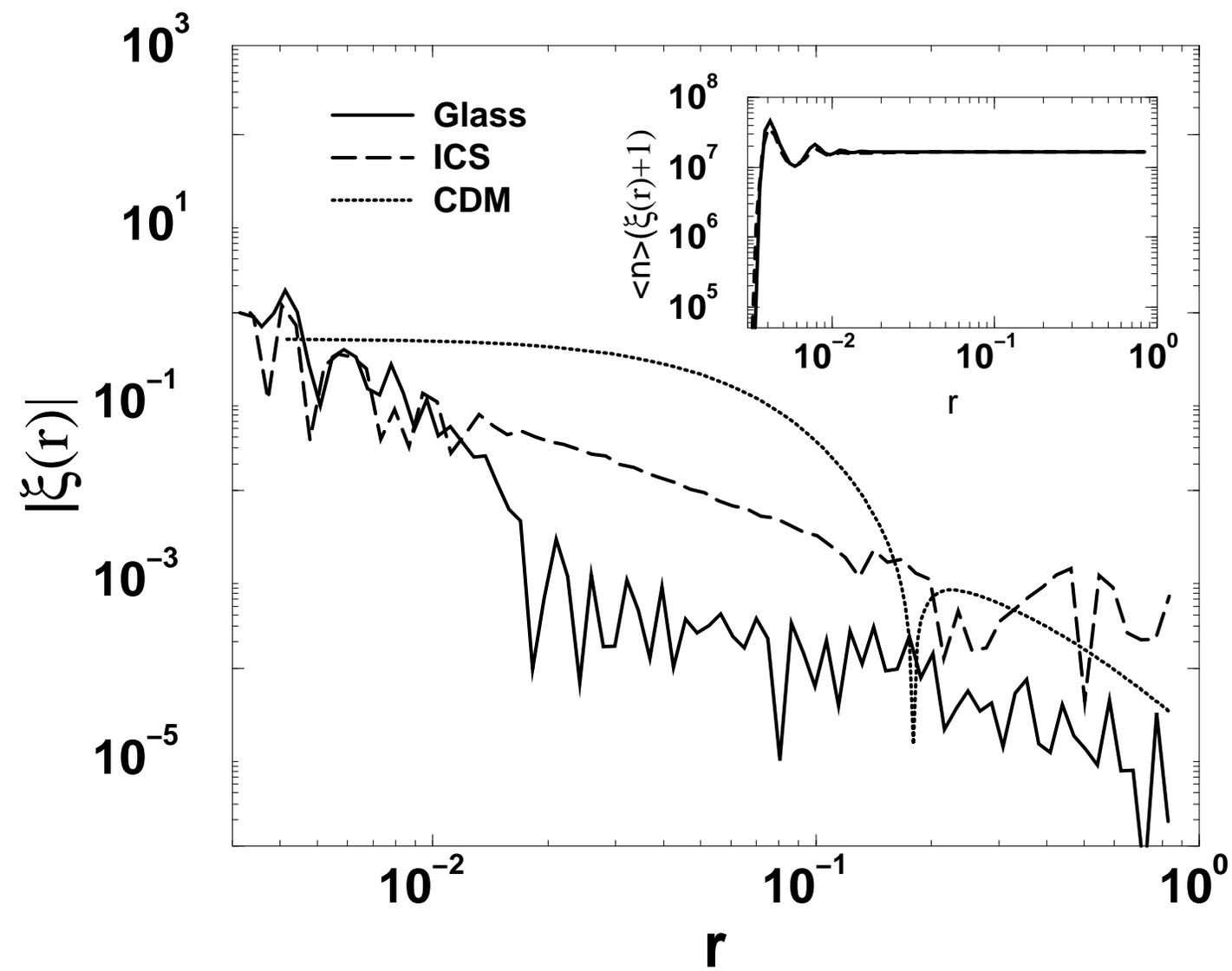
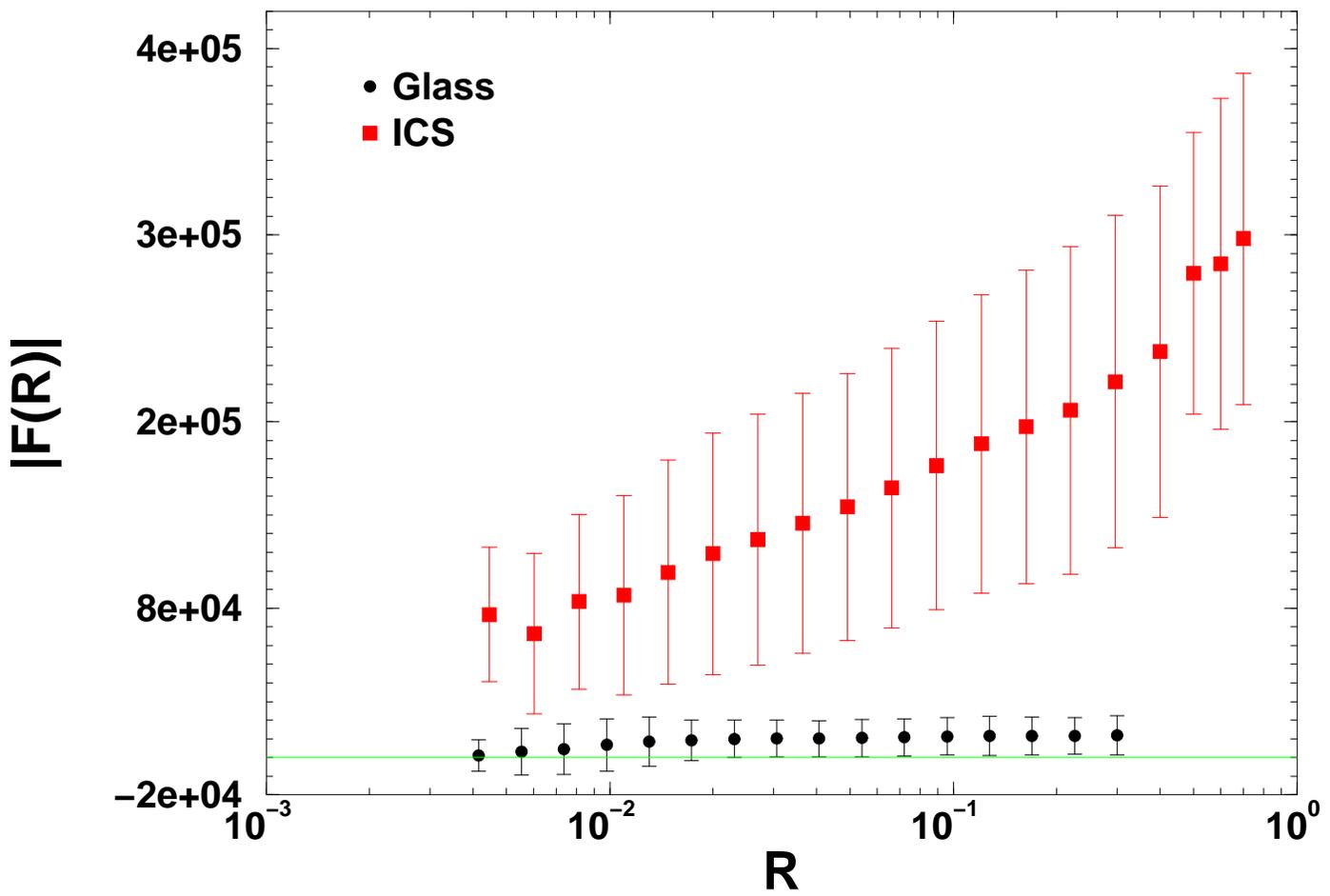
\end{document}